\documentclass[aps,prc,groupedaddress,showpacs,manuscript]{revtex4}

\usepackage{graphicx}

\begin{document}

\title{Deformation and orientation effects in the driving potential of the dinuclear model}

\author {Qingfeng Li$^{1)}$
\email[]{liqf@xinhuanet.com}, Wei Zuo $^{2,3)}$, Wenfei
Li$^{2,3)}$, Nan Wang$^{4)}$, Enguang Zhao$^{1,2,5)}$, Junqing
Li$^{1,2,3)}$\footnote{Corresponding author. E-mail address:
jqli@impcas.ac.cn}, and W. Scheid$^{6)}$}
\address{
1) Institute of Theoretical Physics,
Chinese Academy of Sciences, P. O. Box 2735, Beijing 100080, P. R. China\\
2)Institute of Modern Physics, Chinese Academy of Sciences,
Lanzhou 730000, P.R. China\\
 3) Research Center of Nuclear Theory of National
Laboratory of
Heavy Ion Accelerator of Lanzhou, Lanzhou 730000, P. R. China \\
4) College of Science, Shenzhen University, Shenzhen
518060, P.R. China\\
5)Department of Physics, Tsinghua University, Beijing 100084,
China\\
 6) Institut f\"ur Theoretische Physik,
Justus-Liebig-Universit\"at, Giessen 35392, Germany}


\begin{abstract}
A double-folding method is used to calculate the nuclear and
Coulomb interaction between two deformed nuclei with arbitrary
orientations. A simplified Skryme-type interaction is adopted. The
contributions of nuclear interaction and Coulomb interaction due
to the deformation and orientation of the nuclei are evaluated for
the driving potential used in the description of heavy-ion fusion
reaction. So far there is no satisfactory theory to describe the
evolution of the dynamical nuclear deformation and orientations
during the heavy-ion fusion process. Our results estimated the
magnitude of above effects.
\end{abstract}


\maketitle
\section{Introduction}

The activity of the study on the synthesis of super-heavy elements
is still maintained in both experimentally and theoretically. On
the experimental branch, S. Hofmann et. al \cite{Hof02} in GSI,
Darmstadt, performed experiments on the synthesis and
identification of the nuclei $^{272}111$ and $^{277}112$ in order
to confirm their previous results obtained in the middle of 90s of
last century \cite{Hof95,Hof96}. Furthermore, several additional
decay chains from the reactions $^{64}Ni+^{208}Bi\rightarrow
^{273}111^*$ and $^{70}Zn+^{208}Pb\rightarrow ^{278}112^*$ were
also measured. The joint IUPAC-IUPAP Working Party (IWP) has
confirmed the discovery of element with atomic number 110, which
is named as darmstadtium (Ds), recently the new element with
atomic number 111 has also been proposed  by IWP to be named as
roentgenium (Rg). Experiments on the synthesis of new elements
with atomic numbers 115 as well as 113 in the reaction
$^{243}Am+^{48}Ca$ were carried out at the U400 cyclotron in Dubna
\cite{Oga04}, recently they also reported the results of
excitation-function measurements for the $^{244}Pu+^{48}Ca$
fusion-evaporation reactions for element 114 and the synthesis of
new isotopes of element 116 with the $^{245}Cm+^{48}Ca$ reaction
\cite{Oga042}.

On the theoretical branch, the physics on the more complicated
dynamical process to super-heavy elements has been paid more
attention \cite{Ber01} and investigated by several groups under
different mechanisms, for example, the di-nuclear concept(see the
recent works in \cite{Ada04,LWF03} and the references therein),
the fluctuation-dissipation model \cite{Shen02,Abe03}, the concept
of nucleon collectivization \cite{Zag01,Zag04}, as well as the
macroscopic dynamical model \cite{Swi81,Blo86}.

In the di-nuclear system (DNS) concept
\cite{Vol86,LWF03,Ant93,Ada96,Ada97,Ada04}, the fusion process is
considered as the evolution of a di-nuclear system caused by the
transfer of nucleons from the light nucleus to the heavy one. The
nucleon transfer process is described in Ref.\cite{LWF03} by
solving the master equation numerically. It is found that the
fusion probability of the compound nucleus is very sensitive to
the specific form of the driving potential. In Ref. \cite{LWF03},
the Coulomb interaction potential of deformed nuclei with a
tip-tip orientation is considered. However, spherical nuclei were
adopted in calculating the nuclear interactions since it is
thought that the nuclear interaction does not depend so strongly
on the deformation of the nuclei as the Coulomb interaction due to
the short range characteristics of the nuclear force. Although
some reasonable results, such as the optimal excitation energies,
the residual cross sections of super-heavy compound nuclei, were
obtained for different heavy-ion fusion reactions, the reliability
has to be checked.

Presently a double-folding method is developed to calculate the
nuclear and Coulomb interactions between the two deformed nuclei
with arbitrary orientations. Here we consider the ground state
deformations of the nuclei for all possible combinations of the
DNS of a certain reaction. In principle, the deformed nuclei can
have different relative orientations which supply quite different
conditions for fusion. Some averaging over the orientations of the
nuclei has to be carried out at least in the entrance channel. The
deformation and the orientation evolutions are difficult to be
described, which have not yet been investigated very well by any
model so far. Nevertheless, it is important to bear in mind : what
are the magnitudes that the deformation of nuclei contributes to
the nuclear and Coulomb interactions, respectively, and to explore
how and to which extent the orientations contribute. These
investigations will give a direction for further improvement.

The paper is arranged as follows. In the next section, the
treatment of the nuclear and Coulomb potentials is introduced. We
present the calculated results and the corresponding discussions
in Section III, where the interaction potentials between different
deformed nuclei and their dependence on orientations as well as
the driving potentials used in the DNS model for different
fragmentations in reactions leading to $^{272}Ds$. Finally,
Section IV gives a brief conclusion and outlook.

\section{Treatment of driving potentials for orientated deformed nuclei of DNS }

For a dinuclear system, the local excitation energy is defined as
follows,
\begin{equation}
\epsilon^*=E^*-U(A_1,A_2,R)-\frac{(J-M)^2}{2\mathcal{J}_{rel}}-\frac{M^2}{2\mathcal{J}_{int}}
. \label{excitEn}\end{equation} where $E^*$ is the intrinsic
excitation energy of the dinuclear system converted from the
relative kinetic energy loss, $M$ is the corresponding intrinsic
spin due to the  dissipation of relative angular momentum $J$.
$\mathcal{J}_{rel}$ and $\mathcal{J}_{int}$ are the relative and
intrinsic moments of inertia respectively. $U(A_1,A_2)$ is the
driving potential energy responsible for the nucleon transfer in
the DNS model, and is written down as,

\begin{equation}
U(A_1,A_2,R)=U_{LD+SC}(A_1)+U_{LD+SC}(A_2)-U_{LD+SC}(A_{CN})+U_C(A_1,A_2,R)+U_N(A_1,A_2,R)
, \label{DrivPot}
\end{equation}
where $A_1$, $A_2$, and $A_{CN}$ represent the mass numbers of the
two nuclei and the corresponding compound nucleus, respectively,
we have $A_1+A_2=A_{CN}$. In the DNS model, the driving potential
is normally given as a function of $\eta=(A_1-A_2)/A_{CN}$. The
first three parts of the right hand side of the equation are
calculated from the Liquid-Drop model plus the shell and pairing
corrections \cite{Moe95,Mye65}. $U_C(A_1,A_2,R)$ and
$U_N(A_1,A_2,R)$ are the corresponding Coulomb and nuclear
potential energies between the nuclei and depend on the
fragmentation of the dinuclear system, on the internuclear
distance $R$ and on the orientation and deformation of the nuclei.
They could be calculated by different methods. In the present
work, we calculate them by using the double-folding method

\begin{equation}
U(\mathbf{r}_1-\mathbf{r}_2)=\int{\rho_1(\mathbf{r}_1)\rho_2(\mathbf{r}_2)\upsilon(\mathbf{r}_1-\mathbf{r}_2)d\mathbf{r}_1d\mathbf{r}_2}
 \label{DoubFold}\end{equation}
where $\rho_1(\mathbf{r}_1)$ and $\rho_2(\mathbf{r}_2)$ are the
density distribution of $1$ and $2$ nucleons in a dinuclear
system, $\upsilon(\mathbf{r}_1-\mathbf{r}_2)$ is the corresponding
interaction between the two points. For the nuclear part $U_N$ we
use densities with a smooth falling off at the surface (see later)
and constant densities for the Coulomb interaction. The long-range
Coulomb interaction is not sensitive to the density at the surface
which allows to simplify the calculations. Therefore, we write the
Coulomb interaction as follows,

\begin{figure}
\includegraphics[angle=0,width=0.9\textwidth]{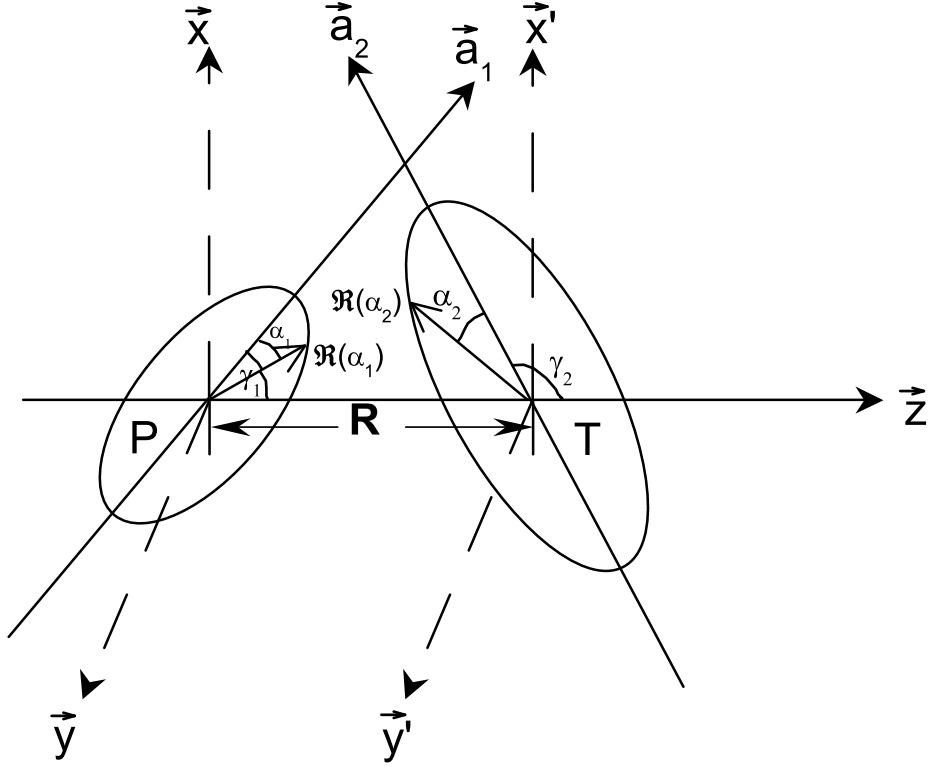}
 \caption{Schematic presentation of the orientation of two axially quadrupolly deformed nuclei.} \label{fig1}
\end{figure}

 \begin{equation}
U_C(R)=\rho_1^0\rho_2^0\int{\frac{d\mathbf{r}_1d\mathbf{r}_2}{|\mathbf{r}_1-\mathbf{r}_2-\mathbf{R}|}},
\label{CoulPot}
\end{equation}
where $\mathbf{R}$ is the vector between the two centers of the
nuclei ("T" and "P") as illustrated in Fig.\ref{fig1}. The charge
densities are set as $\rho_1^0=\frac{Z_1 e}{\Omega_1}$ and
$\rho_2^0=\frac{Z_2 e}{\Omega_2}$ where $Z_{1,2}$ and
$\Omega_{1,2}$ are the proton numbers and the volumes of the two
nuclei, respectively. The symmetry axes ($\vec{a}_1$ and
$\vec{a}_2$) of the two deformed nuclei and the $\vec{z}$-axis are
assumed to be in the same plane. $\gamma_1$ and $\gamma_2$ are the
corresponding angles between the symmetric axes and
$\vec{z}$-axis, {\it i.e.}, which represent the different
orientations of the two nuclei, while $\alpha_1$ and $\alpha_2$
are the angles between arbitrary vectors $\mathbf{r}_{1,2}$ and
the symmetric axes $\vec{a}_1$ and $\vec{a}_2$, respectively. The
distance between the two points is given by
\begin{equation}
|\mathbf{r}_1-\mathbf{r}_2-\mathbf{R}|=\sqrt{(\mathbf{r}_1-\mathbf{r}_2)^2+R^2-2(\mathbf{r}_1-\mathbf{r}_2)\cdot\mathbf{R}}
\label{r12R}\end{equation}

It is easy to find the following relations,
\begin{equation}
(\mathbf{r}_1-\mathbf{r}_2)^2=r_1^2+r_2^2-2r_1r_2(\sin\theta_1\sin\theta_2\cos(\phi_1-\phi_2)+\cos\theta_1\cos\theta_2),
\label{r122}\end{equation}
\begin{equation}
(\mathbf{r}_1-\mathbf{r}_2)\cdot\mathbf{R}=(r_1\cos\theta_1-r_2\cos\theta_2)R.
\label{r12R2}\end{equation} where $\theta_{1,2}$ and $\phi_{1,2}$
are the angles of $\mathbf{r}_{1,2}$ with respect to the
coordinates ($\vec{x}$, $\vec{y}$, $\vec{z}$) and
($\vec{x}\prime$, $\vec{y}\prime$, $\vec{z}\prime$), respectively.

The upper and lower limits of $r_{1,2}$, $\theta_{1,2}$, and
$\phi_{1,2}$ are
\begin{equation}
r_{1,2}:(0,\Re(\alpha_{1,2}));\hspace{0.5cm}\theta_{1,2}:(0,\pi);\hspace{0.5cm}\phi_{1,2}:(0,2\pi),
\label{limits}
\end{equation}
where $\Re(\alpha_{1})$ and $\Re(\alpha_{2})$ describe the nuclear
surface with quadrupole deformations.
\begin{equation}
\Re(\alpha_{i})=R_{0i}(1+\beta_2^iY_{20}(\alpha_{i})).
\label{sphhar}\end{equation} Here $R_{0i}$ are the spherical radii
of the two nuclei which preserve their fixed volumes.
$Y_{20}(\alpha)=(5/4\pi)^{1/2}P_2(\cos\alpha)=(5/4\pi)^{1/2}(3\cos^2\alpha-1)/2$
is spherical harmonics and the axial symmetry is preserved. The
$\beta_2^{i}$ is the quadrupole deformation parameter of
$i$-nucleus taken from Ref.\cite{Moe95}. It is easy to write down
the expressions for $\alpha_1$ and $\alpha_2$ as
\begin{equation}
\cos\alpha_{1}=\hat{\vec{a}}_{1}\cdot\hat{\mathbf{\Re}}(\alpha_1)=\sin\theta_{1}\cos\phi_{1}\sin\gamma_{1}+\cos\theta_{1}\cos\gamma_{1},
\label{cosa1}\end{equation} and
\begin{equation}
\cos\alpha_{2}=\hat{\vec{a}}_{2}\cdot\hat{\mathbf{\Re}}(\alpha_2)=\sin\theta_{2}\cos\phi_{2}\sin\gamma_{2}+\cos\theta_{2}\cos\gamma_{2}.
\label{cosa2}\end{equation}

For the nuclear potential, following the work by Adamian et al.
\cite{Ada96}, we adopt the Skyrme-type interaction without
considering the momentum and spin dependence, in which a
zero-range treatment of the effective interaction
$\delta(\mathbf{r}_1-\mathbf{r}_2)$ is assumed. The nuclear
potential is obtained in the sudden approximation \cite{Ada96},

\begin{eqnarray}
U_N(R)&=&C_0\{\frac{F_{in}-F_{ex}}{\rho_{00}}(\int{\rho_1^2(\mathbf{r})\rho_2(\mathbf{r}-\mathbf{R})d\mathbf{r}}
\nonumber\\
&+&\int{\rho_1(\mathbf{r})\rho_2^2(\mathbf{r}-\mathbf{R})d\mathbf{r}})+\int{\rho_1(\mathbf{r})\rho_2(\mathbf{r}-\mathbf{R})d\mathbf{r}}\}
\label{Un}
\end{eqnarray}
with
\begin{equation}
F_{in,ex}=f_{in,ex}+f'_{in,ex}\frac{N_1-Z_1}{A_1}\frac{N_2-Z_2}{A_2}.
\label{finex}\end{equation} Here $N_{1,2}$ and $Z_{1,2}$ are the
neutron and proton numbers of the two nuclei respectively.
Obviously the isospin effect of the nucleon-nucleon interaction is
considered here though the relative influence is small. The
parameters $C_0=300$ MeV$\cdot$fm$^3$, $f_{in}=0.09$,
$f_{ex}=-2.59$, $f'_{in}=0.42$, $f'_{ex}=0.54$, and
$\rho_{00}=0.17$fm$^{-3}$ are also used in this work. The
functions $\rho_{1}$ and $\rho_2$ are two-parameter Woods-Saxon
density distributions (now we set the center of the "P"-nucleus at
the coordinate origin and $\mathbf{r}_1=\mathbf{r}$)

\begin{equation}
\rho_1(\mathbf{r})=\frac{\rho_{00}}{1+\exp((r-\Re_1(\alpha_1))/a_{\rho_1})}
\label{rho1}
\end{equation}
and
\begin{equation}
\rho_2(\mathbf{r})=\frac{\rho_{00}}{1+\exp((|\mathbf{r}-\mathbf{R}|-\Re_2(\alpha_2))/a_{\rho_2})},
\label{rho2}
\end{equation}
The parameters $a_{\rho_{1}}$ and $a_{\rho_{2}}$ represent the
diffuseness of the two nuclei, respectively. Whereas
$\cos\alpha_1$ is given in Eq. (\ref{cosa1}), we use the following
formula with
$|\mathbf{r}-\mathbf{R}|=\sqrt{r^2+R^2-2rR\cos\theta}$

\begin{eqnarray}
\cos\alpha_2&=&\frac{(\mathbf{r}-\mathbf{R})\cdot{\hat{\vec{a_2}}}}{|\mathbf{r}-\mathbf{R}|}\\
            &=&\frac{r(\sin\theta\cos\phi\sin\gamma_2+\cos\theta\cos\gamma_2)-R\cos\gamma_2}{r^2+R^2-2rR\cos\theta}\nonumber.
\end{eqnarray}

We directly calculate the six- and three-dimensional integrals in
Eqs. (\ref{CoulPot}) and (\ref{Un}) numerically. For Eq.
(\ref{Un}), a truncation parameter $r_{cut}$ for the upper limit
of $r$ is introduced due to the long tails of the nuclear
densities expressed in Eqs. (\ref{rho1}) and (\ref{rho2}). For
each mass asymmetry we calculated the sum of the Coulomb and
nuclear potential energies as a function of the internuclear
distance $R$ and took the potential at the minimum in $R$ which is
shorter than $R_{CB}$ (Coulomb-barrier saddle point) as the
driving potential of the DNS model.

\section{Numerical results}

In this paper, the nuclear and Coulomb interaction for the DNS of
the reaction $^{64}Ni+^{208}Pb\rightarrow ^{272}Ds$ is studied by
taking the nuclear deformations and the corresponding orientations
into account. For simplicity, the diffuseness parameters
$a_{\rho_1}$ and $a_{\rho_2}$ are chosen as
$a_{\rho_1}=a_{\rho_2}=0.6$fm, which is a little bit larger than
those in Ref. \cite{Ada96}. Furthermore, $r_{01}=r_{02}=1.2$fm is
used. The parameter $r_{cut}=25$fm for the radial integration of
the nuclear potential of the deformed nucleus in Eq. (\ref{Un}),
is taken for an adequate precision.

\begin{figure}
\includegraphics[angle=0,width=0.9\textwidth]{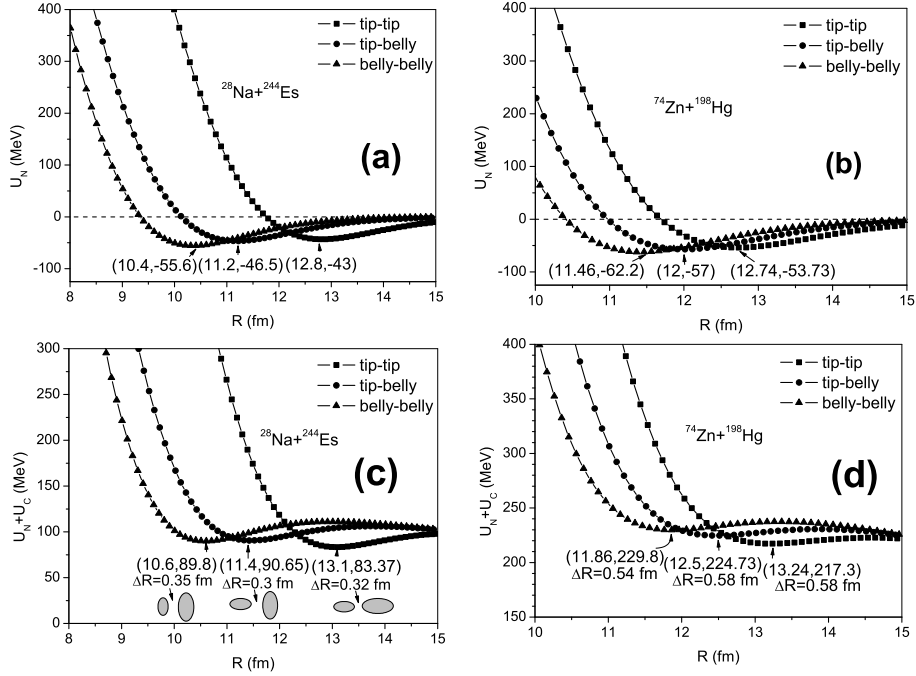}
\caption{The nuclear (in (a) and (b)) and the nuclear+Coulomb
potentials (in (c) and (d)) for two sets of projectile-target
combinations for the same compound nucleus $^{272}Ds$ are shown as
a function of $R$ for different orientations of the two nuclei.}
\label{fig2}
\end{figure}

Fig.\ref{fig2} (a) and (b) show the nuclear interaction potentials
of two sets of projectile-target combinations, namely
$^{28}Na+^{244}Es$ and $^{74}Zn+^{198}Hg$, to form the same
compound nucleus $^{272}Ds$ as a function of distance $R$ between
the centers of the two nuclei. The corresponding nucleus-nucleus
potentials including both the nuclear and Coulomb interactions are
given in Fig.\ref{fig2} (c) and (d). In Fig.\ref{fig2} (a) and
(c), both nuclei are prolately deformed, $^{28}Na$ with
$\beta_2=0.257$ and $^{244}Es$ with $\beta_2=0.234$, respectively,
while in Fig.\ref{fig2} (b) and (d), $^{74}Zn$ is prolate and
$^{198}Hg$ oblate with $\beta_2=0.125$ and $-0.112$, respectively.
The system $^{74}Zn+^{198}Hg$ is more mass-symmetric, {\it i.e.},
it has a smaller $|\eta|$ than the system $^{28}Na+^{244}Es$, and
thus a higher Coulomb potential energy. In each panel, different
orientations for the two systems, {\it i.e.}, tip-tip, tip-belly
and belly-belly orientations are investigated, an illustration is
shown in (c) plot. When both $\beta_2$ values are positive in (a)
and (c), the angles $(\gamma_1,\gamma_2)=(0^0,180^0)$,
$(0^0,90^0)$, and $(90^0,90^0)$ are the corresponding ones for the
tip-tip, tip-belly, and belly-belly cases, respectively, while for
the case of $\beta_2^1>0$ and $\beta_2^2<0$ in cases (b) and (d),
$(\gamma_1,\gamma_2)=(0^0,90^0)$, $(0^0,0^0)$, and $(90^0,0^0)$,
respectively. The two nuclei become more compact with a
belly-belly orientation in contrast to the tip-tip one, and the
minimum of the potential energy for a belly-belly orientation is
at a smaller $R$ than that of the tip-tip case. When the
orientation changes from the tip-tip type to the belly-belly one,
the minima of the nuclear potentials in (a) and (b) behave
differently as those of the total potentials shown in (c) and (d),
{\it i.e.}, the minima of the nuclear interaction go down while
the minima of the total interaction increase. The reason for the
decrease from tip-belly to belly-belly in (c) is that the increase
of the Coulomb interaction energy is smaller than the decrease of
the nuclear interaction energy. Defining a distance between the
surface of the two nuclei, for example, for the tip-tip case,
$\Delta R=R_{min}-(\Re_{1}^{long}+\Re_{2}^{long})$, while for the
belly-belly case, $\Delta
R=R_{min}-(\Re_{1}^{short}+\Re_{2}^{short})$ ( $\Re_{i}^{long,
short}$ represent the long and short axes of the deformed nucleus
$i$, respectively), we find that $\Delta R$ changes a little for
different orientations. When $|\eta|$ decreases from $1$ to $0$,
the value of $\Delta R$ increases due to a larger repulsive
Coulomb force, which can be seen more clearly in Fig.\ref{fig5}.
Therefore, the effect of the mass asymmetry and the orientation of
the DNS on the driving potential can be analyzed from these
results.

\begin{figure}
\includegraphics[angle=0,width=0.9\textwidth]{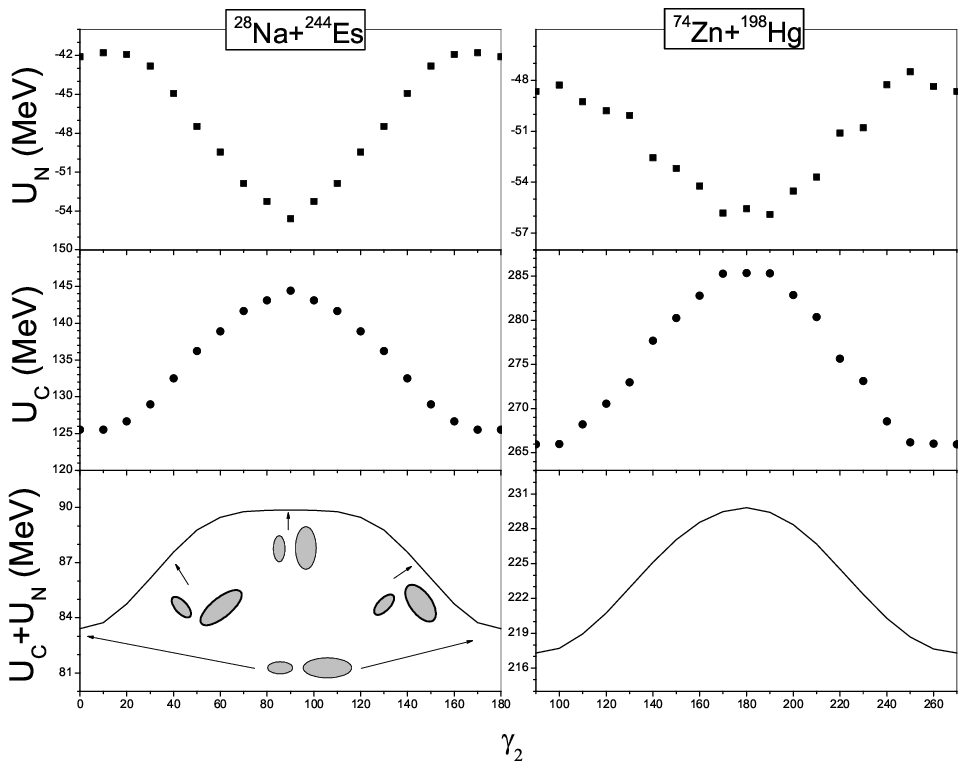}
 \caption{The potentials with $\gamma_1+\gamma_2=180^0$ for $^{28}Na+^{244}Es$ (left panel)
and $\gamma_1+\gamma_2=90^0$ for $^{74}Zn+^{198}Hg$ (right panel).
See text for details.} \label{fig3}
\end{figure}

\begin{figure}
\begin{minipage}[t]{0.48\linewidth}
\centering\includegraphics[width=0.9\textwidth,angle=0.]{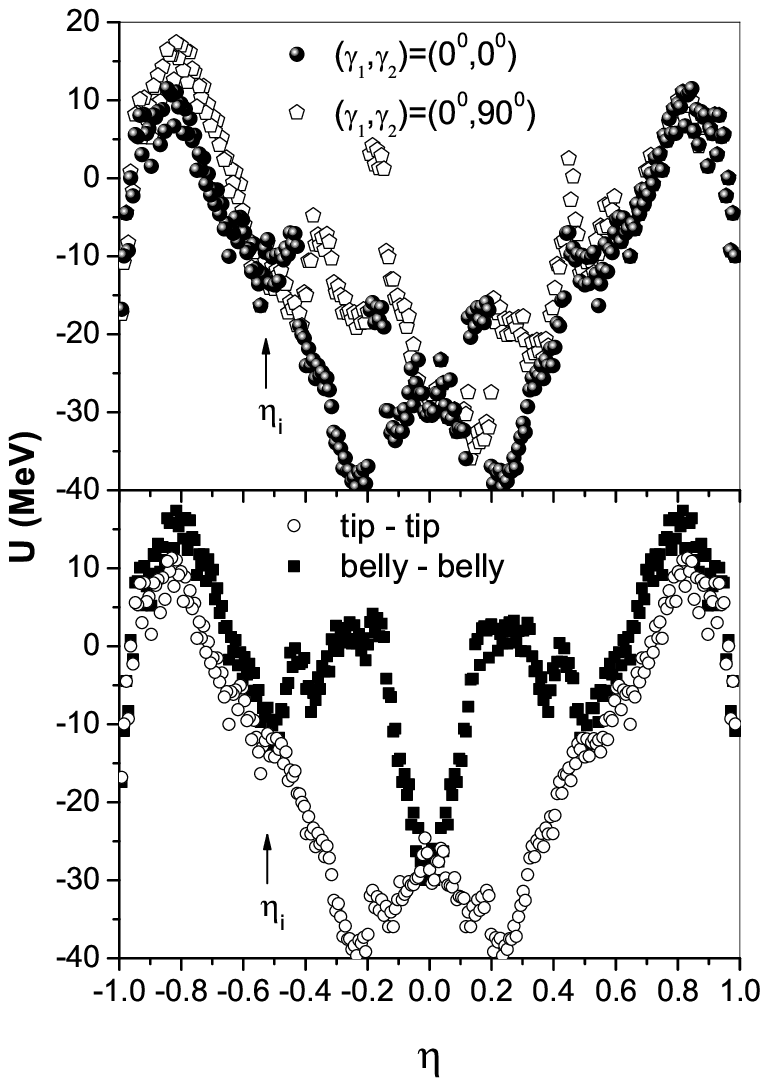}
\caption{The driving potentials with different orientations for
the system $^{272}Ds$. The calculation was started with the
$^{64}Ni+^{208}Pb$ fragmentation(see text).
 } \label{fig4}
\end{minipage}\hspace{5mm}
\begin{minipage}[t]{0.48\linewidth}
\centering\includegraphics*[width=0.9\textwidth,angle=0.]{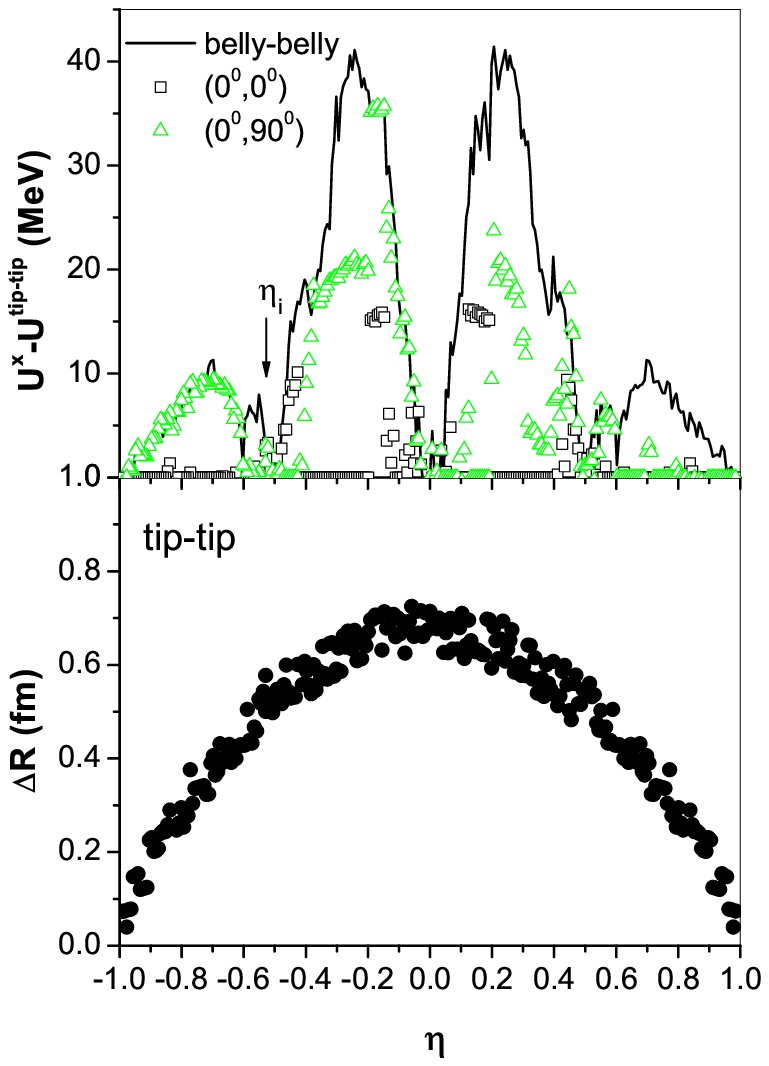}
\caption{(a): the difference between the driving potentials of
tip-tip case and the other cases. (b): $\Delta R$ as a function of
$\eta$ with tip-tip orientation. } \label{fig5}
\end{minipage}
\end{figure}

Fig.\ref{fig3} shows the potentials at the minimum of $U_N+U_C$
illustrated in Fig.\ref{fig2} for the above two combinations as a
function of the orientation, the orientation is chosen in a way
that keeps $\gamma_1+\gamma_2=180^0$ for the system
$^{28}Na+^{244}Es$ and $\gamma_2+\gamma_1=90^0$ for
$^{74}Zn+^{198}Hg$. On the left hand side, $\gamma_1$ goes from
$180^0$ to $0^0$ and $\gamma_2$ from $0^0$ to $180^0$, on the
right hand side, $\gamma_1$ is chosen from $0^0$ to $-180^0$ and
$\gamma_2$ from $90^0$ to $270^0$ in order to obtain similar
trends of the variation of potentials as a function of the
orientation of the two nuclei as on the left hand side, in the two
cases, both of the orientation changes from the tip-tip
orientation to the belly-belly one and finally back to the tip-tip
orientation (the orientation of the nuclei is shown in the
lower-left plot of Fig.\ref{fig3}). With the changing of the
orientations of the two nuclei, the nuclear potentials (upper
panels) form a valley while the Coulomb potentials (middle panels)
attain a peak value for the tip-tip orientation. The summation of
the two contributions shown in the bottom panels is similar in
shape to the Coulomb potential but the change with angle is
gentler.

Fig.\ref{fig4} displays the driving potentials in Eq.
(\ref{DrivPot}) for different orientations. In the upper panel of
the figure we fixed $\gamma_1$ and $\gamma_2$ to $0^0$ or $90^0$,
while in the lower panel, the results for the tip-tip and
belly-belly orientations are shown. The curves in Fig.\ref{fig4}
were calculated by starting with the initial fragmentation
$^{64}Ni+^{208}Pb$ ($\eta_i$) transferring nucleons in steps of
one proton or one neutron by searching for the minimum of
potential energy. Therefore, the potentials are only approximately
symmetric with respect to $\eta=0$ for the tip-tip and belly-belly
cases while for the cases with orientations of $(0^0, 0^0)$ and
$(0^0, 90^0)$ in the upper panel, this symmetry is lost obviously.
From Fig.\ref{fig4}, we find that the driving potential is quite
sensitive to the choice of orientations of the two nuclei. The
driving potential for the tip-tip configuration is smaller than
that for the belly-belly configuration in the whole range of
$\eta$. This result is in disagreement with that obtained in Ref.
\cite{Mis02}, this discrepancy might be associated to the
different consideration of the fusion process of heavy-ions.

To evaluate the difference between the different orientations, we
show the differences between the potential energies of the various
cases with respect to the tip-tip case in the upper half of
Fig.\ref{fig5}, where $U^{belly-belly}-U^{tip-tip}$ is shown with
a line, while the other two cases are shown with different
scattered symbols. The differences are peaked in two regions, one
in $|\eta|<0.5$ and the other in $|\eta|>0.5$. In each region
there exists a large deformation of the nuclei, especially when
$|\eta|$ is $0.1\sim0.4$. However, the detailed deformation of the
two nuclei in each part is different, that is, when $|\eta|>0.5$,
the smaller nucleus is almost spherical while the larger
counterpart is prolately deformed. When $|\eta|<0.5$, prolate and
oblate deformations of the two nuclei occur, for example, for
$\eta=-0.243$, the corresponding configuration is
$^{103}Mo+^{169}Er$ with a couple of large prolate deformation
$(\beta_2^1, \beta_2^2)=(0.358, 0.304)$. When $\eta=-0.169$, the
corresponding $^{113}Pd+^{159}Gd$ consists of a negatively
($-0.25$) deformed $^{113}Pd$ and a positively ($0.28$) deformed
$^{159}Gd$. The separation distance  $\Delta R$ between the
surface of the two nuclei of the DNS is shown in the lower graph
of Fig.\ref{fig5}. Because of the relatively large Coulomb
potential, $\Delta R$ is stretched when the masses of the two
participating nuclei become more equal, which has also been shown
in Fig.\ref{fig2}.


\begin{figure}
\includegraphics[angle=0,width=0.9\textwidth]{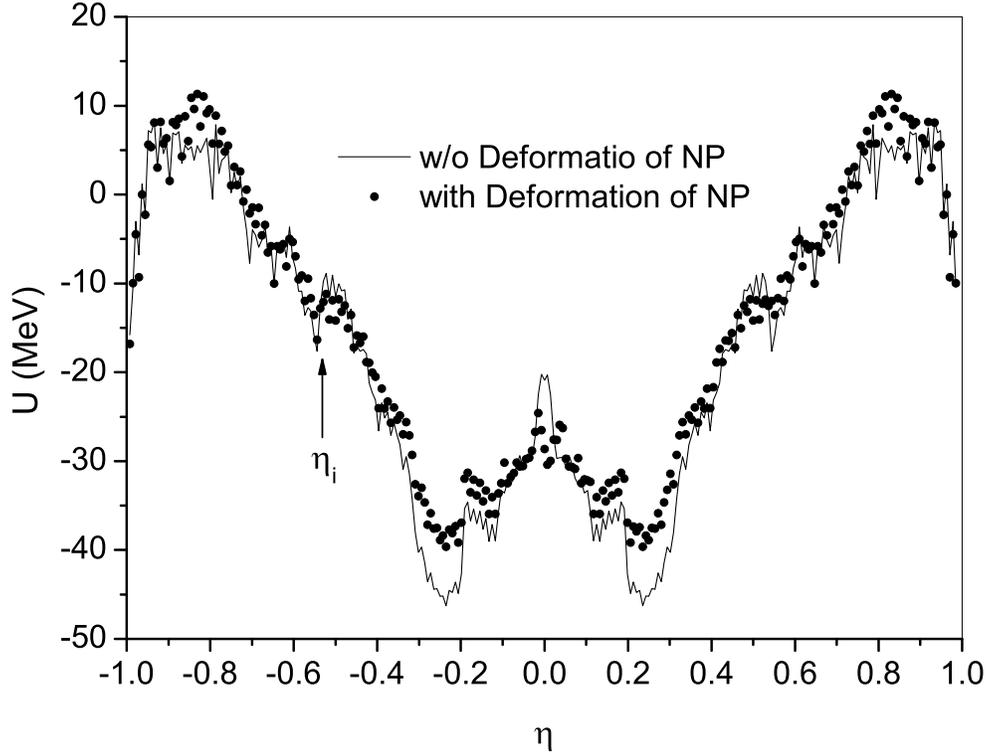}
\caption{The comparison of the driving potential by using the
ground state deformation and the tip-tip orientation for both the
nuclear and Coulomb interactions with previous calculations which
did not taken into account the deformation in the nuclear part of
interaction.} \label{fig6}
\end{figure}

For the di-nuclear system $^{64}Ni+^{208}Pb\rightarrow ^{272}Ds$,
Fig.\ref{fig6} shows the comparison between the present driving
potential  shown by dots and that calculated in Ref. \cite{LWF03}
shown by a fine line for the tip-tip orientation. In the present
calculations, the ground state deformation has been taken into
account for both the nuclear and Coulomb interactions, and in Ref.
\cite{LWF03} a parameterized Morse formula \cite{Ada97} without
considering the deformation of the nuclei has been adopted for the
nuclear potential. We find that the two potentials are basically
very close each other, however, some obvious deviations appear in
the relatively large deformed regions, for example, around
$|\eta|\sim0.2$ and $|\eta|\sim0.8$. It should be pointed out that
a deviation also occurs at $|\eta|\sim0$. After checking the
detailed path of evolution, we find that the configurations of the
DNS in the two cases are different at this point. For the case
with a nuclear interaction of spherical nuclei, the combination
$(^{136}La+^{136}I)$ is preferred, while for the one with that of
deformed nuclei, a more charge-symmetric combination
$(^{136}Ba+^{136}Xe)$ is taken into account. Obviously, the effect
of a large deformation in the deformed region $|\eta|\sim0.2$
changes the final path of the evolution near $\eta=0$.

\section{Conclusion and outlook}
A double-folding method used to calculate the nucleus-nucleus
potential between deformed nuclei is further developed to improve
the driving potential of nuclear fusion in the DNS model. By taken
into account the nuclear deformation in the nuclear interaction
together with the Coulomb interaction, the formalism for
calculating the driving potential of heavy-ion fusion becomes more
reasonable. The deformations and orientations of the interacting
nuclei contributing to the nuclear and Coulomb interactions are
investigated for every fragmentation of the DNS considered. It is
natural that the tip-tip orientation has the lowest interaction
energy, and may be preferred during the nucleon exchange process.
The minimum energies of the nucleus-nucleus interaction along the
distance between the centers of the two nuclei appear at larger
distances when mass-asymmetry $|\eta|$ changes from unit to zero,
which is due to the larger Coulomb force, and is in favor for the
quasi-fission process. So far a dynamical evolution of the
deformation and orientation during the heavy-ion fusion process is
not reasonably treated by the present models to our knowledge, our
results have estimated the effects of the deformation and
orientation of the nuclei, on the driving potential. Hopefully it
will give a direction for the further investigation and
improvement.

In the next step, we will calculate the fusion probability of
various projectile-target combinations with deformed nuclei.
Furthermore, when the distance between the surface of the two
nuclei is elongated the effect of quasi-fission is expected more
pronounced. We will further consider a two-dimensional potential
as functions of the mass asymmetry $\eta$ and the internuclear
distance $R$ in order to investigate the effect of quasi-fission
in the subsequent work.

\section{Acknowledgment}
The authors (Q. Li, W. Zuo, E. Zhao, J. Li, and W. Li) acknowledge
the warm hospitality of the Insitut f\"ur Theoretische Physik,
Universit\"at Giessen, Germany. We are also grateful to Dr. A. D.
Torres for valuable discussions. The work is supported by the
National Natural Science Foundation of China under Grant
No.10175082, 10235020, 10375001, 10311130175; the Major Basic
Research Development Program under Grant No. G2000-0774-07;  the
Knowledge Innovation Project of the Chinese Academy of Sciences
under Grant No. KJCX2-SW-N02; One Hundred Person Project of CSA;
CASK.C. Wong Post-doctors Research Award Fund; the National key
program for Basic Research of the Ministry of Science and
Technology (2001CCB01200, 2002CCB00200) and the financial support
from DFG of Germany.

\end{document}